\documentclass[prl,amsmath,amssymb,twocolumn]{revtex4-1}
\usepackage[normalem]{ulem}
\usepackage{amsmath,amssymb}
\usepackage[bookmarks=true,colorlinks,linkcolor=blue,urlcolor=blue,citecolor=blue]{hyperref}
\usepackage{graphicx,amsmath,amsfonts,dsfont}
\usepackage[usenames]{color}
\usepackage{epstopdf}
\usepackage{verbatim}
\usepackage{amsthm}
\usepackage{enumitem}
\usepackage[caption=false]{subfig} 
\usepackage[mathlines]{lineno} 

\hypersetup{colorlinks,linkcolor=blue,filecolor=green,urlcolor=blue,citecolor=blue}

\newcommand{\be}{\begin{equation}}
\newcommand{\ee}{\end{equation}}

\newcommand{\tr}[1]{\,\mathrm{tr}\left\lbrace  #1 \right\rbrace}
\newcommand{\ket}[1]{\left\vert#1\right\rangle}
\newcommand{\bra}[1]{\left\langle#1\right\vert}

\newcommand{\average}[1]{\left\langle#1\right\rangle}

\newcommand{\syse}{\end{array}\right.}

\begin{document}

\title{Quantum Critical Scaling  under Periodic Driving}

\author{Salvatore Lorenzo$^{1,2,8}$}
\author{Jamir Marino$^{3}$}
\author{Francesco Plastina$^{4,5}$}
\author{G. Massimo Palma$^{1,6}$}
\author{Tony J. G. Apollaro$^{1,2,6,7}$}

\affiliation{$^1$ Dipartimento di Fisica e Chimica, Universit$\grave{a}$  degli Studi di Palermo, via Archirafi 36, I-90123 Palermo, Italy}
\affiliation{$^2$ Quantum Technology Lab, Dipartimento di Fisica, Universita` degli Studi di Milano, 20133 Milano, Italy}
 \affiliation{$^3$Institute of Theoretical Physics, University of Cologne, D-50937 Cologne, Germany}
 \affiliation{$^4$Dip. Fisica, Universit$\grave{a}$ della Calabria, 87036, Arcavacata di Rende (CS), Italy}
\affiliation{$^5$INFN - Gruppo collegato di Cosenza, Cosenza Italy}
\affiliation{$^6$NEST, Istituto Nanoscienze-CNR, Pisa, Italy}
\affiliation{$^7$Centre for Theoretical Atomic, Molecular and Optical Physics,
School of Mathematics and Physics, Queen's University Belfast, Belfast BT7 1NN, United Kingdom}
\affiliation{$^8$INFN, Sezione  di  Milano,  I-20133  Milano,  Italy}
\affiliation{Correspondence and requests for materials should be addressed to T.J.G.A. (email tony.apollaro@gmail.com)}


\begin{abstract}
{\bf Universality is key to the theory of phase transitions,
stating that the equilibrium properties of observables near a
phase transition can be classified according to few critical
exponents.
These exponents rule an universal scaling behaviour that witnesses the irrelevance of the model's microscopic details at criticality.
Here we discuss the persistence of such a scaling
in a one-dimensional quantum  Ising  model under sinusoidal modulation in time of its transverse magnetic field.
We show that scaling of various quantities (concurrence, entanglement entropy, magnetic and fidelity susceptibility)
endures up to a
stroboscopic time $\tau_{bd}$, proportional to the  size of the system. This behaviour is explained by noticing that the low-energy modes, responsible for the scaling properties, are resilient to the absorption of energy. Our results suggest that relevant features of the universality  do hold also when the system is brought out-of-equilibrium by a periodic driving.}

\end{abstract}


\date{\today}

\maketitle

A paradigm of  phase transitions is the concept of universality,
i.e., the insensitivity to microscopic details at the  critical
point of many particle systems at equilibrium. Universality allows
to classify phase transitions according to critical exponents,
which govern the scaling of several quantities close to the
critical point. A  quantum many body system at zero temperature
can encounter a phase transition driven by quantum fluctuations
when some of its control parameters are tuned to a critical value,
which in the simplest case separates an ordered from a disordered
phase \cite{Sachdev}.
As an hallmark of these quantum phase transitions (QPT), and in
analogy with classical (temperature-driven) ones, physical
observables exhibit scaling properties near such quantum critical
point (QCP), with their leading behaviour depending only on few
universal critical exponents \cite{FSS72, domb1983phase}.
The insensitivity to microscopic physics and the emergence of
universal critical exponents is in turn a consequence of the
absence of any typical length scale in the system, as the
correlation length diverges at criticality.

Temperature, however,  plays a detrimental role in QPT, as it sets a thermal
de Broglie length above which long-range correlations are
suppressed \cite{Sachdev}. This aspect renders the search for an analogue
quantum critical behaviour in non-equilibrium (NEQ) many body
systems a challenging task:
an external agent pumping energy into the system~\cite{DallaTorre2012, Sieberer13, marino, Marino2016bis, Russomanno6}  induces an effective temperature
delimiting the NEQ critical scaling region up to a characteristic  'thermal' length.

Among the rich varieties  of NEQ drivings, perturbations periodic
in time~\cite{Grifoni}  constitute a promising platform to engineer
hopping in optical lattices \cite{Weiss05, Burton13,
Meinert16}, artificial gauge fields~\cite{ Atala13,
Goldman14}, topological phases of matter \cite{Kitagawa11,
Jotzu14, Bloch15, Refael11, Levin13, RigolChern}  as well
as to induce dynamical localisation effects \cite{Prosen1,
Prosen2, Alessio13, Ponte15}.
In general, a non-adiabatic perturbation heats a
system~\cite{PhysRevLett.113.260601}, and in absence of a bath
dissipating the injected energy~\cite{Refael, Ponte15} or peculiar
conditions -- such as integrability \cite{LazaridesDas14, Angelo}  or
disorder-induced localisation effects \cite{CC, LazMBL}, a fully
mixed,  infinite temperature state will be eventually approached
in the long-time dynamics.
However, it has been shown that at intermediate time scales  novel
interesting  effects can still  be  observed in an isolated
periodically driven, ergodic system,  such as the onset of NEQ
long-lived metastable states~\cite{DeR15, Demler15,
Weid}.
\\
In this study, we consider the \emph{resilience of  critical scaling
under periodic driving}.
Specifically, we show the {\it{robustness}} of critical scaling
exponents when a time-periodic modulation is super-imposed on the
transverse magnetic field of a Quantum Ising model~\cite{Angelo,
Angelo2, Russomanno15} prepared in its critical ground state.
Indeed, a number of quantities, evaluated on the time dependent
out-of-equilibrium state of the periodically driven system,
follows the same scaling~\cite{domb1983phase} behaviour proper of
the equilibrium QCP, even though the state itself is very far from
the critical ground state.
This behaviour persists up to a stroboscopic time scale,
$\tau_{bd}$, where  scaling breaks down,
%
thus setting a condition for the observation of quantum critical
scaling in periodically driven many body systems.

\section{Results}
\subsection{Periodically driven Ising model}
We investigate the 1D quantum $XY$-model, driven by a periodic
transverse magnetic field:
\begin{equation}\label{E.XYHam}
\hat{H}(t){=}-\sum_{i=1}^{N}\left(\frac{1+\gamma}{2}\hat{\sigma}_i^x\hat{\sigma}_{i+1}^x{+}
\frac{1-\gamma}{2}\hat{\sigma}_i^y\hat{\sigma}_{i+1}^y{-}h(t)\hat{\sigma}_i^z\right)~,
\end{equation}
where $\hat{\sigma}^{\alpha}$ ($\alpha{=}x,y,z$) are the Pauli
matrices, $h(t){=}h{+}\Delta h \sin\left(\omega t\right)$ is the
harmonically modulated  transverse field, and $\gamma$ the
anisotropy parameter. For $\gamma\in\left(0,1\right]$, this model
belongs to the Ising universality class and it exhibits a  second order
QPT with a critical point located at $h=h_c=1$,  separating a ferromagnetic phase from a
paramagnetic  one.
%
The $XY$-model with a static field is diagonalised by standard
Jordan-Wigner (JW) and Bogolyubov transformations
~\cite{PhysRevA.2.1075, Sachdev}, enabling Eq. \eqref{E.XYHam}
(with $\Delta h=0$) to be  re-written as a free fermion
Hamiltonian
\begin{equation}\label{E.Hamdia}
\hat{H}=\sum_k
\varepsilon_k\left(2\hat{\gamma}_k^{\dagger}\hat{\gamma}_k-1\right)~.
\end{equation}
Here $\varepsilon_k{=}\sqrt{\left(h-\cos k\right)^2+\left(\gamma
\sin k\right)^2}$ is the energy of the Bogolyubov quasiparticle
with momentum $k$, and annihilation operator
$\hat{\gamma}_k{=}u_k\hat{c}_k{+}v_k\hat{c}_{-k}^{\dagger}$, the
$c_k$'s being JW spinless fermion operators labelled by the
momentum $k$.
The ground state of Eq.~\ref{E.Hamdia} can be written in the BCS form
\begin{equation}\label{E.GS}
\ket{GS}\equiv\ket{\Psi(t{=}0)}{=}\prod_{k>0}\ket{\psi(0)}_k{=}\prod_{k>0}\left(u_k{+}v_k \hat{c}_k^{\dagger}\hat{c}_{-k}^{\dagger}\right)\ket{0}~,
\end{equation}
with $\ket{0}$ representing the vacuum of the JW fermions
($\hat{c}_k \ket{0} = 0 \, , \forall k$).
The  ground state \eqref{E.GS} at the critical point ($h=h_c$) is
the initial state for the periodic drive considered in this study.

Floquet analysis is a valuable tool
to deal with time-periodic Hamiltonians,
$\hat{H}(t)=\hat{H}(t+nT)$, as it allows to reduce the {\it{stroboscopic}} time evolution,
i.e. at integer steps $n$ of the period $T$, to a dynamics
generated by a {\it{time-independent}} effective
Hamiltonian~\cite{Grifoni}.
%
%
Indeed, by exploiting the periodicity of  $\hat{H}(t)$, the
(stroboscopic) unitary time evolution operator $\hat{U}(t)$, at times
$t=n T$, can be written as a discrete-time quantum map
$\hat{U}(nT)=\left[\hat{U}(T)\right]^n$, where the effective
(Floquet) Hamiltonian, $\hat{H}_F$, is hence defined~\cite{LazaridesDas14} by
$\hat{U}(T)=e^{-i T \hat{H}_F}$. In the periodically driven XY
model, the Floquet operator can be expressed as a product of
operators acting in each two-dimensional $k$-subspace, spanned by
the vacuum and by the state with a pairs of JW fermions
$\{\ket{0_k 0_{-k}},\ket{1_k 1_{-k}}\}$, namely
$\hat{U}(T)=\prod_{k>0}\hat{U}_k(T)$.
Accordingly, after $n$ periods of the sinusoidal drive,  the
initial critical state \eqref{E.GS} evolves into
$\ket{\Psi(nT)}=\hat{U}(nT)\ket{\Psi(0)}=\prod_{k>0}\hat{U}_k(nT)\ket{\psi(0)}_k$
(for further details see Suppl. Mat. of Ref.~\cite{Angelo}).

$\ket{\Psi(nT)}$ will be the stroboscopic state where we test the
persistence of finite size scaling (FSS) behaviour, the
characteristic trait of criticality, in several quantities and for
a number of different driving conditions.

\subsection{Critical scaling under periodic drive}
In the following, we provide evidence that the scaling behaviour
of a number of physical quantities, which should strictly hold at
the equilibrium QCP only, persists in fact also under periodic
driving.
In particular, we consider both local -- in the real lattice space, quantities (e.g.,
nearest-neighbor concurrence and local transverse magnetic
susceptibility) and non-local ones (e.g., entanglement
entropy and fidelity susceptibility).

Remarkably, scaling behaviour at equilibrium has been found also
for quantities that are not observables in the strict quantum
mechanical sense, such as entanglement. Indeed, the
concurrence~\cite{PhysRevLett.80.2245}, a measure of bi-partite
entanglement for two qubits (see Methods), has been shown to exhibit FSS at
equilibrium QCP, as first demonstrated for the quantum Ising
model~\cite{Osterloh2002} and later illustrated in other
systems~\cite{PhysRevA.71.060304,dicke1,dicke2,
PhysRevA.68.042330,PhysRevB.81.064418}. The scaling of concurrence
was explored in details, as it bridges QPT with quantum
information theory~\cite{pmid24807201, PhysRevLett.115.236601} and
quantum thermodynamics,  because of its close connection with the
notions of irreversible work~\cite{PhysRevB.93.201106} and
ergotropy~\cite{francica}.

%
%

In the quantum Ising (QI) model at the equilibrium critical point
~\cite{Osterloh2002}, the derivative of the concurrence between
neighboring spins with respect to the transverse field, $h$,
displays a logarithmic singularity at $h=h_c$. An FSS analysis at
equilibrium for the concurrence shows data collapse for different
system sizes, consistent with the universal critical exponent
$\nu=1$, which also governs the divergence of the correlation
length of the order parameter in the QI model.

\begin{figure} [ht!]
 \centering
    \includegraphics[width=0.5\textwidth]{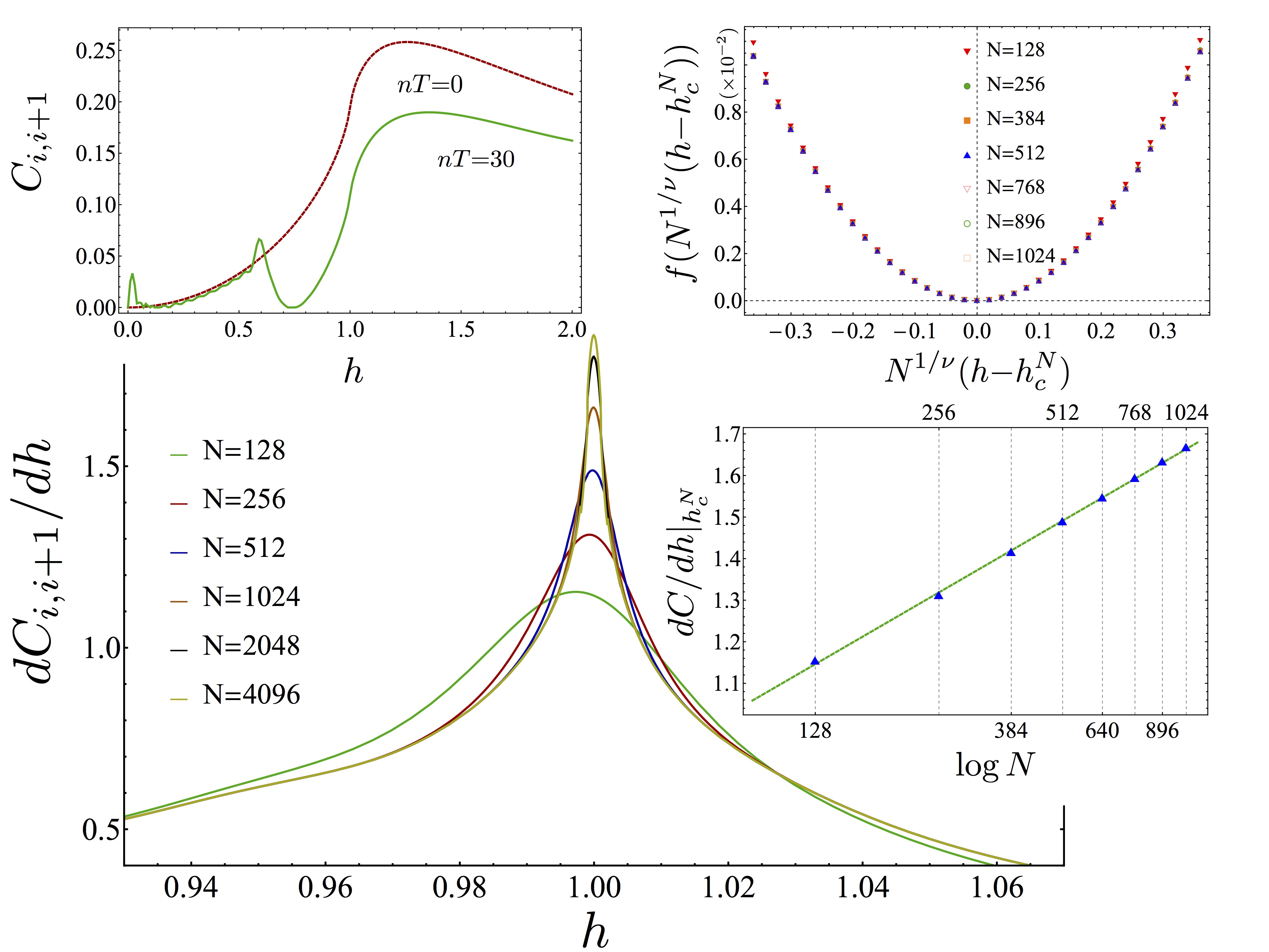}
    \caption{    {\bf{Finite-size scaling of the concurrence.}}
Main (central) plot: $\frac{ dC_{i,i+1}(N)}{dh}$ vs. initial
magnetic field $h$ around the QPT point for different system's
sizes $N$. A logarithmic divergent behaviour is found close to $h=1$, as inferred from the right lower inset, where the maximum of the peaks of $\frac{ dC_{i,i+1}(N)}{dh}$ vs $N$ are reported in a semilogarithmic plot. Upper
Left  inset: Nearest-neighbor Concurrence  $C_{i,i+1}(N)$ vs.
initial magnetic field $h$ at equilibrium, for an infinite chain
(red-dotted line) and after $n=30$ cycles for $N=128, 256,
512,1024,2048,4096$, (green continuous lines: different lines are
not distinguishable on this scale). Upper right inset: Data
collapse for FSS at the logarithmic divergence, for the same $N$
as listed above, after $n=30$ cycles, according to the Ansatz
$1{-}\exp\left(\frac{d C(N)}{dh}{-}\frac{d
C(N)}{dh}\big|_{h=h_c^N}\right)=f\left(N^{\frac{1}{\nu}}\left(h-h_c^N\right)\right)$,
with $\nu=1$.
The driving protocol is  $h(t)=1{+}0.1\sin 2
\pi t$, and we took $\gamma=1$ (cfr. Eq.~\ref{E.XYHam}). The chosen value of $n=30$ is within the breakdown time of the shortest chain here considered (see following section). For  $n<30$, although the numerical values of  $\frac{ dC_{i,i+1}(N)}{dh}$  (and $C_{i,i+1}(N)$) change, the FSS data collapse and the logarithmic divergence, upper right and left inset, respectively, is attainable with the same critical exponent. }
     \label{F_C12}
\end{figure}

Interestingly, we find that the same scaling property still
holds for a periodically driven, out of equilibrium system.
The nearest-neighbor concurrence $C_{i,i+1}(N)$ for different system sizes $N$ is reported in the upper left inset of Fig.~\ref{F_C12} both at equilibrium, $nT=0$, and after time $nT=30$ of the
driving $h(t){=}h{+}0.1 \sin\left(2\pi t\right)$. Notice that, notwithstanding the system is driven out-of-equilibrium, the qualitative features of $C_{i,i+1}(N)$ around the critical point $h_c$ are preserved.
Indeed, in the main plot of Fig.~\ref{F_C12}, we report the derivative of
the stroboscopic nearest-neighbor concurrence,
$\frac{dC_{i,i+1}(N)}{dh}$, in the neighborhood of $h_c$ for
different system sizes $N$ and after $n=t/T=30$ cycles of the
driving $h(t){=}h_c^N{+}0.1 \sin\left(2\pi t\right)$, where
$h_c^N$ is the pseudocritical point,
where $h_c^N$ is the pseudocritical point, which moves towards $h_c$ as $h_c-N^{-2} (\log N +const)$. It is interesting to note that the logarithmic correction to the shift exponent $\lambda=2$ is shared also by the equilibrium FSS behaviour of the half-chain entanglement entropy~\cite{IgloiLinJSM08}.
With increasing system's size $N$, a logarithmic divergence in the
concurrence builds up at $h=h_c^N$, namely $\frac{
dC_{i,i+1}(N)}{dh}\Bigr|_{h=h_c^N}\propto \log N$ (see the lower right inset of Fig.
\ref{F_C12}).
%
Performing the FSS analysis for logarithmic
divergences~\cite{domb1983phase}, we obtain  data collapse once the
scaling exponent  is set to its equilibrium value $\nu=1$, as
reported in the upper right inset of Fig.~\ref{F_C12}.
FFS is attained with the same value of the critical exponent $\nu$
even after changing driving amplitude, frequency, anisotropy
parameter as well as number of cycles, provided the conditions
outlined in the following subsection are fulfilled. This
highlights that the role played by universality stretches well
beyond the ground state properties, significantly affecting the
system even under periodic driving.

In the following we will substantiate further our claim about the
persistence of FSS in periodically driven critical systems by
considering additional quantities.
For this purpose, we have also considered  the scaling relation for
the transverse magnetic susceptibility of the $XY$-model,
$\chi_z^{(N)}(h){=}\frac{1}{N}\frac{d\average{\hat{M}^z}}{dh}{=}\frac{d\average{\hat{\sigma}^z}}{dh}$.
At equilibrium, it exhibits a scaling behaviour with critical
exponent $\alpha=0$~\cite{mccoy1973two}, implying a logarithmic
divergence akin to the one encountered for the concurrence.
Such logarithmic divergence is preserved also under driving,
and, in analogy with the scaling Ansatz for the concurrence, data
collapse is obtained for different system's sizes, implying that
even after the system is brought out-of-equilibrium by periodic
driving, the scaling exponents keep their equilibrium values
$\alpha=0$ and $\nu=1$.
\begin{figure} [ht!]
\centering
    \includegraphics[width=0.45\textwidth]{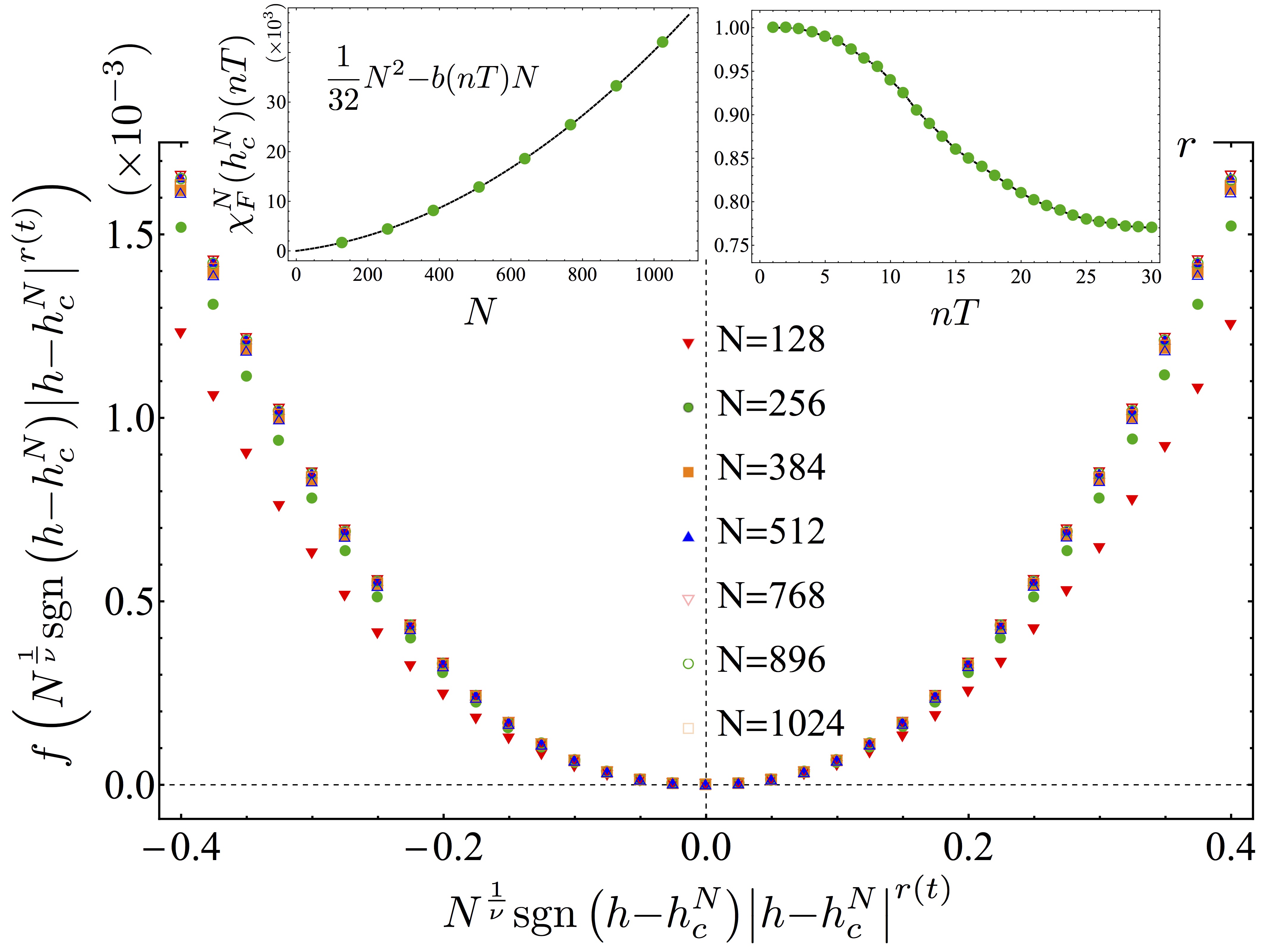}
\caption{{\bf{Finite-size scaling of the Fidelity
susceptibility.}} Main plot: Data collapse of the Fidelity
susceptibility according to the Ansatz in Eq.~\ref{E.FSSFS} after
$n=15$ periods, with exponents $r(15 T)=0.86$ and
$\nu=1$. Left inset: The maximum of the Fidelity susceptibility
attained at the pseudocritical point $\chi_F^N(h_c^N)$ after
$n=15$ driving periods. The plot shows an algebraic divergence
with the system size. Right inset: Values of the time-dependent
exponent $r(nT)$ that allows for the FSS data collapse.}
     \label{Fid_sus_sca}
\end{figure}
So far, we have considered single and two-site quantities.
Still, physical quantities having support on a larger part
of the system, e.g., the entanglement entropy, or even genuinely
global, such as the Fidelity susceptibility, exhibit FSS at
equilibrium in the critical QI model~\cite{IgloiLinJSM08, book_Dutta}. The
former following a logarithmic and the latter an algebraic
divergence, respectively.
As for the Entanglement Entropy of the half chain $S_{N/2}$,
defined by the von Neumann entropy $S_{N/2}=-\tr{\hat{\rho}_{N/2}
\log_2  \hat{\rho}_{N/2}}$, we find that the FSS relation
$S_N(h){-}S_{N/2}(h_c^N){=}f\left(N^{\frac{1}{\nu}}\left(h-h_c^N\right)\right)$,
based on the logarithmic law of the entanglement entropy at
criticality, derived in Ref.~\cite{IgloiLinJSM08}, holds as well
under periodic driving up to times $t_{N/2}{=}\frac{N}{2
v_{max}}$, where $v_{max}$ is the maximum group velocity of the
Floquet quasiparticles~\cite{2016arXiv160303579A, Nag1, Nag2} (see Methods). Indeed
$t_{N/2}$ gives a time after which the quasi-particles have left
the half chain and a volume law for the Entanglement Entropy is
attained~\cite{2016arXiv160303579A, 1742-5468-2016-7-073101,Cincio2007}.

Let us now turn our attention to the Fidelity susceptibility (FS).
The ground state FS is defined by $\left|\bra{GS(h)}GS(h+\delta
h)\rangle\right|$ and it depends on three length scales; namely,
the system size $N$, the correlation length $\xi\sim \left |
h-h_c\right|^{-\nu}$ and the length scale associated to the
parameter $\delta h$, $\xi_{\delta h}\sim \left|\delta
h\right|^{-\nu}$. If $\xi_{\delta{h}}$ is the largest length
scale, it is meaningful to consider the FS~\cite{DeG,Damski1}, defined by
$\chi_F^N(h){=}-\frac{\partial^2 F}{\partial(\delta h)^2}$, whose
scaling behaviour will be dictated by the other two length scales
(for a detailed analysis of the use FS in transverse field spin
models see the book by Dutta {\it{et al.}} in
Ref.~\cite{book_Dutta}). For the Ising model at equilibrium, it
has been shown
that, at criticality, where $\xi\gg N$, the FS exhibits a maximum
whose height scales algebraically with the system size as
$\chi_F^N(h_c^N){=}\frac{1}{32}\left(N^2-N\right)$. Far from
criticality, on the other hand, where $\xi\ll N$, the scaling is
extensive~\cite{Damski1, Damski2}. In the following we will
investigate the former limit as the driving is around criticality.
For the analysis of the stroboscopic Fidelity susceptibility in
the limit where $\xi\ll N$, see the Section Methods.
Notice that, contrary
to the quantities considered before, $\chi^N_F$ does not scale
logarithmically with $N$. Notwithstanding, FSS is still
attainable, provided a new, time-dependent exponent is introduced,
which takes into account the fact that the algebraic scaling gets
modified. At equilibrium, one considers the susceptibility of the
ground state fidelity, $F^N(h){=}\left|\bra{GS(h)}GS(h+\delta
h)\rangle\right|$ for $\delta h\rightarrow 0$, which, by
definition, is time-independent and can be related also to the irreversible work in an infinitesimal quench protocol~\cite{Paganelli1}. On the other hand, in the
presence of the driving, we will consider the susceptibility of
the following expression for the Fidelity:
$F^N(h)(nT){=}\left|\bra{GS(h)}\hat{U}^{\dagger}(nT)\hat{U}'(nT)
\ket{GS(h+\delta h)}\right|$, where $\hat{U}$ and $\hat{U}'$
differ as they correspond to driving around $h$ and $h+\delta h$,
respectively. $F^N(h)(t=nT)$ reduces to the ground state Fidelity
at $t=0$.
In Fig.~\ref{Fid_sus_sca}, we report our results for the fidelity
susceptibility, $\chi_F^N$, obtained by taking $\delta h=10^{-5}$
and the same driving parameters as in Fig.~\ref{F_C12}.
Interestingly enough, we find that the algebraic divergence is
preserved, according to the law
$\chi_F^N(h_c^N)(nT){=}\frac{1}{32}N^2{-}b(nT)N$, where $b(nT)$
($b(0){=}\frac{1}{32}$) is a monotonically increasing function
(see left inset in Fig.~\ref{Fid_sus_sca}). This behaviour can be
qualitatively explained by noticing that the driving ultimately will
invalidate the FSS behaviour (hence a linear scaling with $N$
is retrieved) at later times (see following subsection). As a consequence, we modify the FSS
Ansatz~\cite{Fidelity_GU_IJMPB}, by introducing a time-dependent
exponent $r(nT)$ ($r(0)=1)$,
\begin{equation}\label{E.FSSFS}\begin{split}
&\frac{\chi_F^N(h_c^N)(nT){-}\chi_F^N(h)(nT)}{\chi_F^N(h)(nT)}{=}\\
&=f\left(N^{\frac{1}{\nu}} \mbox{sgn }\left(h{-}
h_c^N\right)\left|h{-} h_c^N\right|^{r(nT)}\right)~,
\end{split}\end{equation}
where $\mbox{sgn}(\cdot)$ is the sign function. In the main plot
of Fig.~\ref{Fid_sus_sca} we show an instance of the data collapse
obtained by means of Eq.~\ref{E.FSSFS} after $n=15$ periods, where
$r(nT)=0.86$. By choosing a different number of periods, the
exponent $r(nT)$ decreases following the curve reported in the
right inset of Fig.~\ref{Fid_sus_sca}. It is worth mentioning that
our $r(0)=1$ coincides with the equilibrium value, which is given by
$r(0)=\frac{d_a^c-d_a}{\nu}$, where $d_a^c=2$ and $d_a=1$ are the
critical adiabatic dimension and the adiabatic dimension,
respectively~\cite{Fidelity_GU_IJMPB}. The $d$'s cannot be easily
defined out-of-equilibrium, and, therefore, we relied on numerical
tools to extract the best collapse exponent $r(nT)$. Nevertheless,
notice that the critical exponent $\nu$ equals 1 at all times
where the FSS holds.

Notably, all the quantities whose FSS we have reported so far for
the Ising model ($\gamma=1$ in Eq.~\ref{E.XYHam}), maintain this
scaling behaviour, with the same critical exponents, in the whole
XY universality class $0<\gamma\leq 1$.

\begin{figure}[ht!]
        \includegraphics[width=0.45\textwidth]{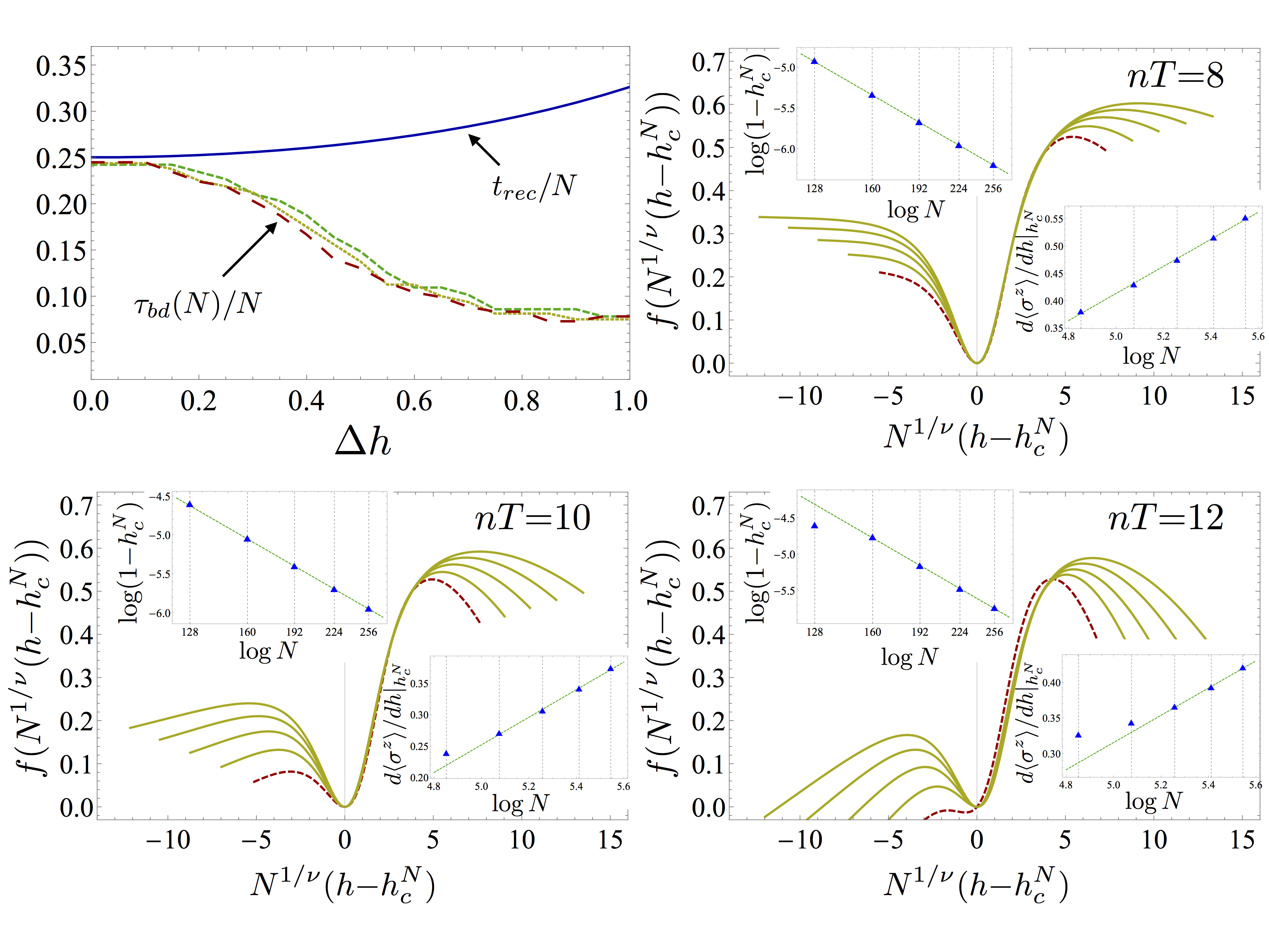}
\caption{{\bf{Breakdown of finite-size scaling.}} (Upper left)
Maximum number of cycles for which FSS holds for the
$z$-susceptibility, with different system sizes,
$N=\{128,160,192\}$, as a function of the applied amplitude for a
driving frequency $\omega=4$ in the Ising model ($\gamma=1$ in
Eq.~\ref{E.XYHam}). Notice the relation $N_2 \tau_{bd}(N_1)\simeq
N_1 \tau_{bd}(N_2)$, where $N_1$ and $N_2$ are different system
sizes, indicating that the breakdown time scales linearly with
$N$. The continuous blue  line shows the recurrence time $t_{rec}$
and one sees that, for driving amplitudes $\Delta h<0.1$,
$\tau_{bd}\simeq t_{rec}$. In the other three panels, we report an
instance of the breakdown of FSS for driving parameters $\omega=4$
and $\Delta h=0.75$ in the Ising model, taken after $n=8,10$, and
$12$ driving periods, respectively, with
$N=\{128,160,192,224,256\}$. We see that the curves corresponding
to lower $N$ depart from the FSS Ansatz more and more with
increasing $n$, while data collapse is still achieved for the
other sizes. The insets in each figure report the algebraic fit of
the pseudocritical point $h_c^N$ (upper left inset) and the
logarithmic divergence of the maximum of the magnetic
susceptibility $\chi_z^{(N)}(h_c^N)$, showing that, at the breakdown time $\tau_{bd}(N)$,  only the points corresponding to the lower $N$'s do not satisfy the respective scaling relation.}
    \label{F.time_panel}
\end{figure}

\subsection{Breakdown time of finite-size scaling}
The FSS under periodic driving holds up to a characteristic time
$\tau_{bd}(N)$, which we dub {\it{breakdown time}}, and which
depends on the driving parameters. Remarkably, we obtain  for
$\omega>4$ and $\Delta h \ll h_c$, a breakdown time comparable to
the recurrence time $\tau_{bd}(N)\simeq t_{rec}$ (where
$t_{rec}{=}N/(2 v_{max})$), while for smaller frequencies, the
breakdown time occurs before the onset of recurrences
($\tau_{bd}<t_{rec}$).
Hence, for a large frequency and small amplitude driving, the
breakdown of critical FSS could be arbitrarily delayed by taking
systems of a large-enough-size. For the high-frequency limit, see
also a recent paper by Gritsev and Polkovnikov~\cite{Gritsev17},
where it has been shown that the Floquet Hamiltonian of a
step-like periodically driven Ising model shares the same critical
properties of the time-independent Ising model based on the
structure of the Onsager algebra and the self-duality (in the
limit of $\omega\rightarrow \infty$, the system freezes in its
initial state and FSS is attainable for every $n$ and $\delta h$,
but this would fall us back into the equilibrium scenario.)

On the other hand, the breakdown of FSS at times smaller than
$t_{rec}$ that occurs for small frequencies ($\omega <4$) is
mostly due to the fact that systems with small sizes loose their
scaling properties. As an instance of such a behaviour, in
Fig.~\ref{F.time_panel} we display $\tau_{bd}(N)$, as extracted
from the transverse magnetisation susceptibility. In the upper
left panel, we show that the breakdown time is comparable to the
recurrence time for small driving amplitudes; conversely, by
increasing $\Delta h$, the number of cycles for which FSS holds
reduces significantly,
as can be expected by observing that already in the first few cycles a considerable amount of energy is absorbed by the system. In the same figure, we report three
instances of scaling analysis performed at increasing number of
cycles. FSS holds for all cases but the one corresponding to the
system of smaller size, so that all of the curves collapse near
$h_c$, but one.

\begin{figure}[ht!]
        \includegraphics[width=0.45\textwidth]{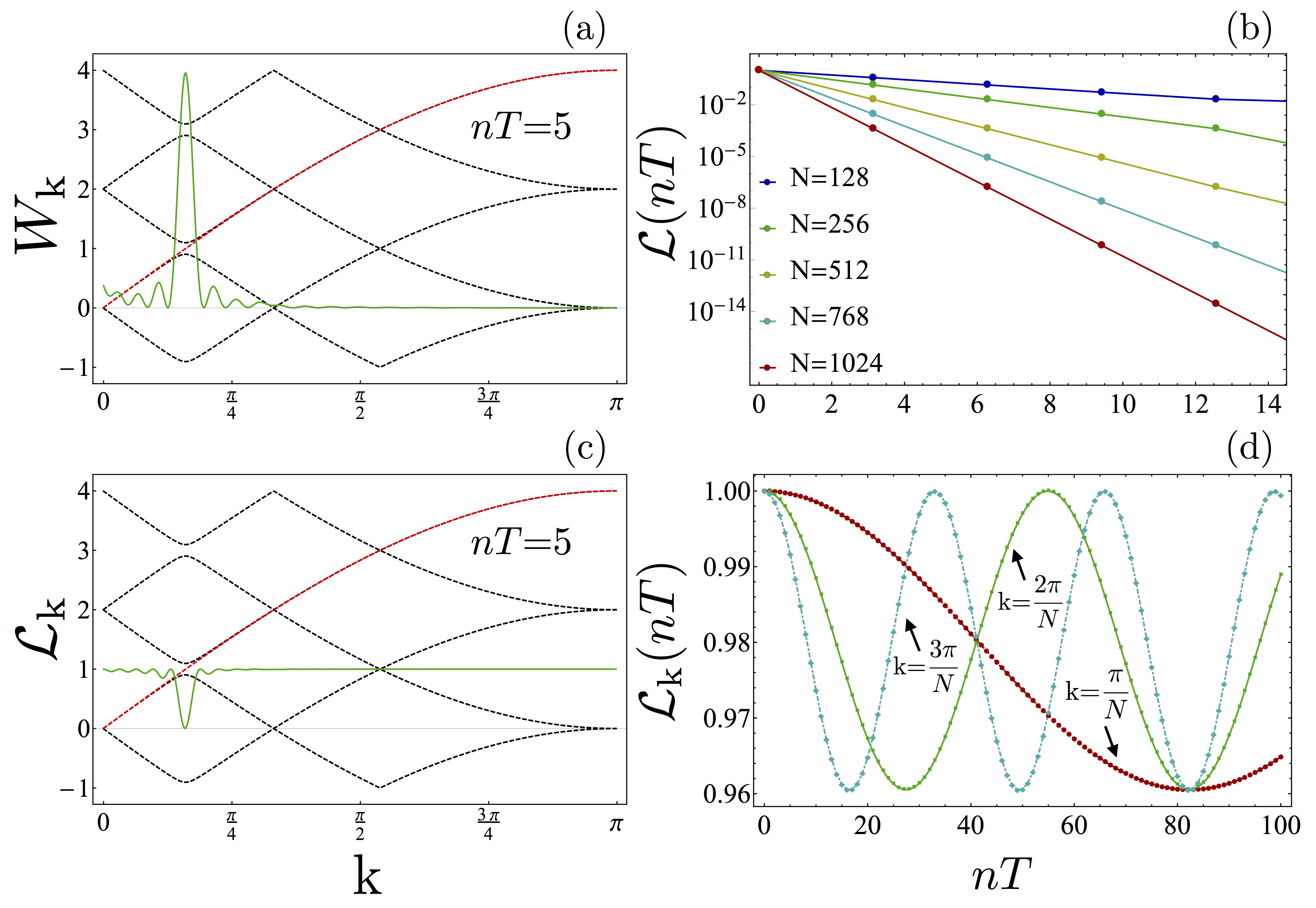}
\caption{{\bf{Loschmidt echo}} (a) Work performed by the driving
agent after $t=5 T$, resolved in $k$:  modes
near the quasi degeneracy in the Floquet spectrum absorb much more
than the others. (b) Loschmidt echo evaluated at the stroboscopic
times, showing an exponential decay. (c) Momentum resolved
Loschmidt echo evaluated at the fixed time $t=5 T$, showing that
modes near the quasi-degeneracy give the dominant contribution to
the decay of $\mathcal{L}$. (d) Stroboscopic time evolution of the
$k$-resolved Loschmidt echo, for the longer wavelength modes,
showing an oscillatory behavior. All the plot are drawn for a
driving with $\omega=2$ and $\Delta h=0.1$. The two left panels (a
and c) also report the Floquet spectrum (dashed black lines) and
the single-particle energies of the unperturbed Hamiltonian Eq.~\ref{E.Hamdia} (red line) on the background.}
    \label{Fidelity}
\end{figure}

We stress that the time scales for which FSS persists are well
beyond the very short transient where the system remains in the
close neighborhood of its initial critical equilibrium state.
Indeed, by evaluating the Loschmidt Echo (see Methods),
$\mathcal{L}(nT)=\left|\bra{\Psi(nT)}GS\rangle\right|$, which
gives the probability amplitude to find the system in a state
close to the initial critical ground state~\cite{SharmaEPL}, we find  -- already
after a few cycles-- that $\mathcal{L}(nT)$ has become negligibly
small. This is explicitly shown in Fig. \ref{Fidelity} (b), where
an exponential decay of $\mathcal{L}$ is reported. A closer look,
however, shows that the decay of the Loschmidt echo is essentially
due to those mode $k$ for which a quasi-degeneracy occurs in the
Floquet energies \cite{Angelo}. These almost degenerate modes are
also those responsible for energy absorption from the driving, see
Fig. \ref{Fidelity}.

Actually, the Loschmidt echo can be decomposed as a product of
contributions from the different modes, $\mathcal{L}= \prod_k
\mathcal{L}_k$ (see the methods section). Such a momentum resolved
Loschmidt echo is shown in Fig. \ref{Fidelity} (c), where it is
superimposed to the dispersion relation of the Floquet quasi
energies (given by the black dashed lines on the background). One
can see that the minimum of $\mathcal{L}_k$ is found where an
`inter-band quasi-degeneracy' occurs.  Moreover, this
`quasi-resonance' precisely corresponds to the peak of the energy
absorbed from the driving, as shown in Fig. \ref{Fidelity} (a).
There, the amount of work performed by the driving agent is
reported for each mode $k$ (see Methods). Therefore, we can conclude that the
decay of the Loschmidt echo is due to these specific modes, where
a Floquet resonance occurs, and which are brought significantly
out of equilibrium by the absorption of energy from the driving.


On the other hand, one expects FSS to be essentially a feature of
the long wavelength (and low energy) modes. These modes are much
less affected by the driving as they absorb much less energy than
those close to the quasi degeneracy. Correspondingly, their
contribution to the overall decay of the Loschmidt echo is very
small.  The time dependence of the $k$-resolved Loschmidt echo for
such small-$k$-modes is reported in Fig. \ref{Fidelity} (d), where
we see that $\mathcal{L}_k(nT)$ periodically oscillates in time.
For each k, the period of such oscillations is determined by the k-eigenvalue of $\hat{H}_F$ (i.e. $\hbar /  \mu_k$ see Methods), which becomes larger and larger with
increasing $N$.

It turns out that the breakdown of FSS occurs close to the time
where a minimum is found for the $\mathcal{L}_k$ corresponding to
the smallest $k$. Therefore, $\tau_{bd}(N) \approx \hbar /(2
\mu_k)$. After this time, in fact, it not possible anymore
to obtain the collapse of curves corresponding to physical
observables evaluated for different $N$'s. This is due to the fact
that systems with different sizes behave asynchronously. This
means that, although the $\mathcal{L}_k$ corresponding to the
smallest $k$ goes back to unity after a period, this revival
occurs at a time that explicitly depends on the system size. As a
result, since the scaling is a comparison of physical quantities
for different $N$ at the same time, it does not hold anymore
because systems with different sizes have $k$-resolved Loschmidt
echoes that are `out of phase' from each other.

Finally, let us comment on the fact that for low-$\omega$
drivings, the FSS behaviour is lost already after a single cycle
because the low-$k$ modes absorb more energy from the drive as the
Floquet resonances move towards them, (see Methods).

\section{Discussion}\label{S.Discussion}

In summary, we have shown that the scaling behaviour proper of an
equilibrium quantum critical point retains its validity also when
the system is brought out-of-equilibrium via a periodic drive.
Finite-size scaling of both local and global quantities,
exhibiting logarithmic as well as algebraic scaling with the
system size, has been performed. We have shown that the
equilibrium critical exponents are robust against the periodic
perturbation up to times when the stroboscopic state is far from
the quantum critical state, suggesting that the features of
universality manifest themselves also under strong periodically
driven settings. In addition, our claims are within reach of experimental verification, as out-of-equilibrium quantum Ising dynamics is currently under active investigation via a great variety of different physical systems, ranging from degenerate Bose gases in a optical lattice~\cite{Ising_exp1, Ising_exp2} to Rydberg atoms~\cite{Ising_exp3}.
Our results may find applications also in the emerging field of out-of-equilibrium quantum thermodynamics, where, recently~\cite{Campisi2016}, quantum Otto engines, having as working substance a many-body system at the verge of criticality, have been suggested to be able to attain the Carnot efficiency at finite power because of the validity of the FSS relations.

%
%

It would be interesting to study in the future whether the same scenario holds
for non-integrable systems hosting a quantum phase transition, as,
for instance, the one dimensional Bose-Hubbard model, where, however,
 the system is eventually driven into an infinite temperature state, and therefore the persistence of critical scaling is expected only
in a temporal window delimited by the thermalization time of the system.
A different scenario, on the other hand, could emerge for interacting integrable models, such as the antiferromagnetic XXZ Hamiltonian, where the N\'{e}el and the XY phase are separated by a second-order QPT and thermalisation is prevented by the integrability of the model.

\section{Methods}\label{S.Methods}

\subsection{Concurrence}
The concurrence, $C(\rho)$, is a measure a bipartite qubit
entanglement, which can be straightforwardly computed for any
two-spin-$\frac{1}{2}$ density matrix
$\rho$~\cite{PhysRevLett.80.2245}. In particular, we apply it to
evaluate the entanglement between two spins of the chain, residing
at sites $i$ and $j$. In this case, the two spin reduced density
matrix, $\hat{\rho}_{i,j}$, is obtained by computing the partial
trace over all but the $i$-th and $j$-th spin degrees of freedom
of either i) the ground state $\ket{GS}$, or ii) the
stroboscopical state $\ket{\Psi(nT)}$. Given the state, $C_{i,j}$
can be evaluated, once the state $\hat{\rho}_{i,j}$ is expressed
in the logical basis of the eigenstates of the $\hat{\sigma}^z$
operator, via the relation
$C=\max\left[0,\sqrt{\lambda_1}-\sum_{n=2}^4\sqrt{\lambda_n}\right]$,
where the $\lambda_n$ are the eigenvalues, in decreasing order, of
$\hat{\tilde{\rho}}=\left(\hat{\sigma}^y\otimes\hat{\sigma}^y\right)\hat{\rho}^*\left(\hat{\sigma}^y\otimes\hat{\sigma}^y\right)$.

\subsection{Loschmidt echo}
The Loschmidt echo is defined as
$$\mathcal{L}(nT) = \left | \left \langle \Psi(t=0) | \Psi(nT) \right
\rangle\right | \, .$$ Using the explicit form of the initial and
stroboscopic state, namely \begin{eqnarray*} \ket{\Psi(0)} &=&
\prod_{k>0} \left ( u_k(0) + v_k(0) \hat{c}^{\dag}_k \hat{c}^{\dag}_{-k}
\right ) \ket 0 \, , \\
\ket{\Psi(nT)} &=& \prod_{k>0} \left ( u_k(nT) + v_k(nT)
\hat{c}^{\dag}_k \hat{c}^{\dag}_{-k} \right ) \ket 0 \, , \end{eqnarray*} we
obtain $\mathcal{L} (nT) = \prod_k \mathcal{L}_k(nT)$, where the
$k$-resolved Loschmidt echo is given by
$$\mathcal{L}_k(nT) = \left |u_k(0) u_k(nT) + v_k(0) v_k(nT) \right | \, .$$
For an extensive analysis of the Loschmidt echo in periodically driven systems see Ref.~\cite{SharmaEPL}.

\subsection{Work}
The (average) work performed up to time $t$ by driving the system
is given by the difference between the average instantaneous
energy of the system and its initial value given by the ground
state energy. At the discrete time instants $t=nT$, we have
$$W (nT)= \bra{\Psi(nT)} \hat{H} (0) \ket{\Psi(nT)} - E_{GS} \, ,$$
where we used the fact that $\hat{H}(nT) = \hat{H}(0)$.

Using the explicit expression of $\hat{H}$ in terms of the Bogoliubov
fermions, Eq. \ref{E.Hamdia}, we have that the constant terms
cancel out and that the work naturally decomposes into the sum of
contributions arising from each mode $k$,
$$W(nT) = \sum_k W_k(nT)\, ,$$
where $$W_k(nT) = 2 \varepsilon_k \, \bra{\Psi(nT)}
\hat{\gamma}^{\dag}_k \hat{\gamma}_k \ket{\Psi(nT)} \, .$$ Notice that, since
the Hamiltonian undergoes a periodic driving, and we are
evaluating the work at an integer number of periods, the average
work coincides in our case with both the so called irreversible
work and the inner friction \cite{PhysRevLett.113.260601}, so that
it can be used to describe also the amount of irreversibility
brought into the system.

\subsection{Floquet spectrum}

The time-independent effective Hamiltonian, dubbed {\textit{Floquet}} Hamiltonian $\hat{H}_F$ and corresponding to the time-dependent Ising Hamiltonian in Eq.~\ref{E.XYHam} of the main text,  can be expressed in quadratic form~\cite{LazaridesDas14} as,
\begin{equation}\label{E.SM.HFlo}
\hat{H}_F=\sum_{k>0}  \hat{h}_{kF}=\sum_{k>0}\mu^+_k \left(\hat{\mu}^+_k\right)^{\dagger} \hat{\mu}^+_k+\mu^-_k\left(\hat{\mu}^-_k\right)^{\dagger} \hat{\mu}^-_k~,
\end{equation}
where $\left\{\mu^{\pm}_k,\ket{\mu^{\pm}_k}\equiv
\left(\hat{\mu}^{\pm}_k\right)^{\dagger}\ket{0}\right\}$ are,
respectively, the positive and negative Floquet eigenvalues and
eigenvectors of the Floquet Hamiltonian $ \hat{h}_{kF}$  for the
mode $k$. The evolution operator $\hat{U}_k(T)=e^{-i T
\hat{h}_{kF}}$ is determined by the solution of the Bogoliubov-de
Gennes equations
\begin{equation}\label{E.SM.BdG}
i \begin{pmatrix}\dot{u}_k(t)\\\dot{v}_k(t)\end{pmatrix}=\begin{pmatrix}\cos k - h(t)& \sin k\\ \sin k & -\cos k + h(t)\end{pmatrix}\begin{pmatrix}u_k(t)\\v_k(t)\end{pmatrix}~,
\end{equation}
for each mode $k$ with the initial condition $\left\{v_k(0),u_k(0)\right\} = \{0,1\}$ after one period $t=T$.
Because of the periodicity of the Hamiltonian in Eq.~\ref{E.XYHam} of the main text, the Floquet eigenvectors are defined up to a periodic phase, corresponding to a shift of the eigenenergies of an integer multiple of the driving frequency
$\omega$,
$\mu^{\pm}_k \rightarrow \mu^{\pm^{(l)}}_k =  \mu^{\pm}_k+l \omega$.
 The latter symmetry brings to the definition of the Brillouin zones $BZ(l)=-\left[(l-1)\frac{\omega}{2},l \frac{\omega}{2}\right)\cup \left[(l-1)\frac{\omega}{2},l \frac{\omega}{2}\right)$ as those reported in Fig.~\ref{Fidelity} in the main text.
 As a consequence, resonances can occur both within the same band and between different bands, dubbed intra-band and inter-band resonances, respectively, in the main text.

Finally, as the stroboscopic evolution of the initial state is given by $\ket{\Psi(nT)}=e^{-i T \hat{H}_F}\ket{\Psi(0}$, the dynamics is governed by the Floquet quasiparticles energies $\mu_k$ and hence the maximum velocity quasiparticles can spread out is given by
$v_{max}=\underset{k}{\max} \frac{d \mu_k}{d k}$,
that is the maximum of the ''group velocity'' as given by the "dispersion relation" of
$ \hat{H}_F$.

\subsection{Extensive scaling of the Fidelity susceptibility far from criticality}

\begin{figure}[ht!]
\begin{center}
\includegraphics[width=0.45\textwidth]{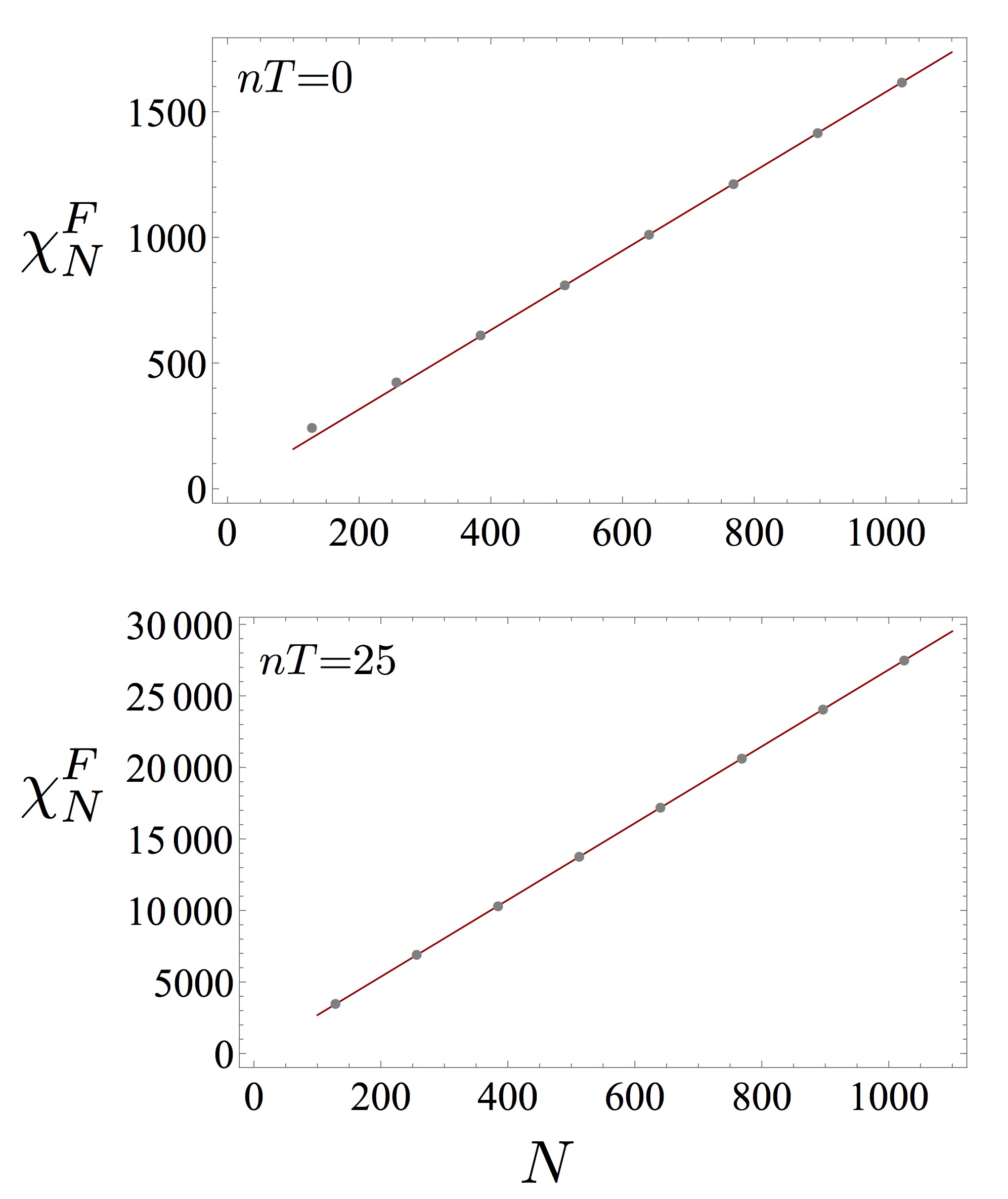}
\caption{{\bf{Non-critical linear scaling of the fidelity
susceptibility.}}(upper plot) Linear scaling of the fidelity
susceptibility at equilibrium in the region where $N\gg \xi$.
(lower plot) The linear scaling is preserved also after $n=25$
cycles of the periodic drive. The dynamical parameters in both
plots are: $h(0)=0.95$ and $\delta h=10^{-5}$ in the fidelity
$F^N(h)$ of the main text, and, in addition, $\omega=2 \pi$, and
$\Delta h=0.1$ for the stroboscopic fidelity $F^N(h)(nT)$.}
\label{F.FSlinN}
\end{center}
\end{figure}

In the main text we have investigated the fidelity susceptibility
(FS) in the limit where the order parameter correlation length
$\xi\sim \left| h-h_c\right|^{-\nu}$ dominates over the system
size $N$, i.e., $\xi\gg N$. In such a regime, the FS scales as
$N^2$ at equilibrium and   FS finite-size scaling has been
reported in Fig.~\ref{Fid_sus_sca} of the main text. On the other
hand, if  $N\gg \xi$, the FS scales at equilibrium linearly with
$N$. Here we will show that such a linear scaling is preserved
also under periodic drive. In Fig.\ref{F.FSlinN} we report the
fidelity susceptibility both at equilibrium and after $n=25$
cycles of the driving.

\begin{figure}[ht!]
\begin{center}
\includegraphics[width=0.45\textwidth]{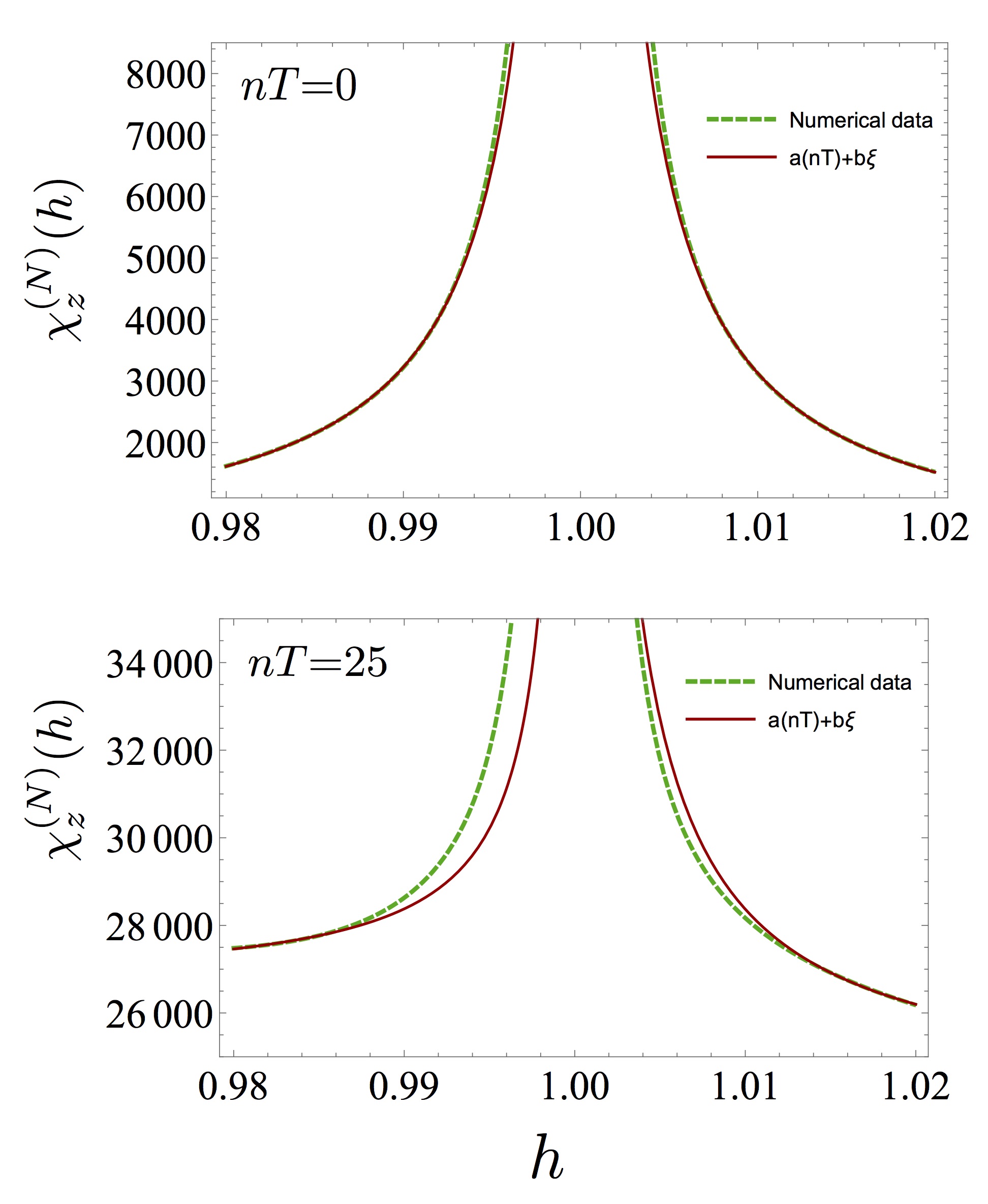}
\caption{{\bf{Scaling of the fidelity
susceptibility with the magnetic field.}} Fidelity susceptibility $\chi_z^{(N)}(h)$ in the limit $N\gg \xi$ in the paramagnetic ($h>1)$
and ferromagnetic phase ($h<1$) at equilibrium (upper plot) and after $n=25$ cycles (lower plot). The fitting curve $\chi_z^{(N)}(h)=a(nT)+b\, \xi$, where $\xi = \left(\ln h\right)^{-1}$ accurately overlaps with the numerical results at $nT=0$, as expected away from criticality. At $n=25$, the scaling of $\chi_z^{(N)}(h)$ as $\left(\ln h\right)^{-1}$ is still visible, although the range of validity 
has decreased to points further away from criticality than in the equilibrium case.
Both plots are reported for $N=1024$, $\delta h=10^{-5}$, $\omega=2 \pi$, and
$\Delta h=0.1$ in the driving protocol.}
\label{F.FSchi}
\end{center}
\end{figure}

Furthermore, by fixing $N$, in the limit $N\gg \xi\sim \left(\ln h\right)^{-1}$, i.e., away from criticality,
it is known that at equilibrium the fidelity susceptibility $\chi_z^{(N)}(h)$ scales as $\xi$~\cite{book_Dutta}. In Fig.~\ref{F.FSchi} we report, 
for $N=1024$, $\chi_z^{(N)}(h)$ as a function of $h$ for drivings both in the ferromagnetic and the paramagnetic phase.
By considering values of $h$ such that the correlation length fulfills $\xi \ll N$,
we notice that the scaling of $\chi_z^{(N)}(h)$ as $\xi$ is preserved also under periodic drive, although the range of validity of such a scaling shrinks by increasing the number of periods $n$.
Nevertheless, a fitting curve of the type  $\chi_z^{(N)}(h)=a(nT)+b\, \xi$ overlaps with our numerical results still after $n=25$ periods of the driving for values of $h$ further away from criticality.
Notice also that the fitting curve has a time-dependent coefficient, $a(nT)$, 
that plays the same r\^{o}le of $b(nT)$ for the FSS behaviour of $\chi_z^{(N)}(h)$ at criticality, see Fig.~\ref{Fid_sus_sca}.

\subsection{Low-$\omega$ drivings}

\begin{figure}[ht!]
\begin{center}
\includegraphics[width=0.45\textwidth]{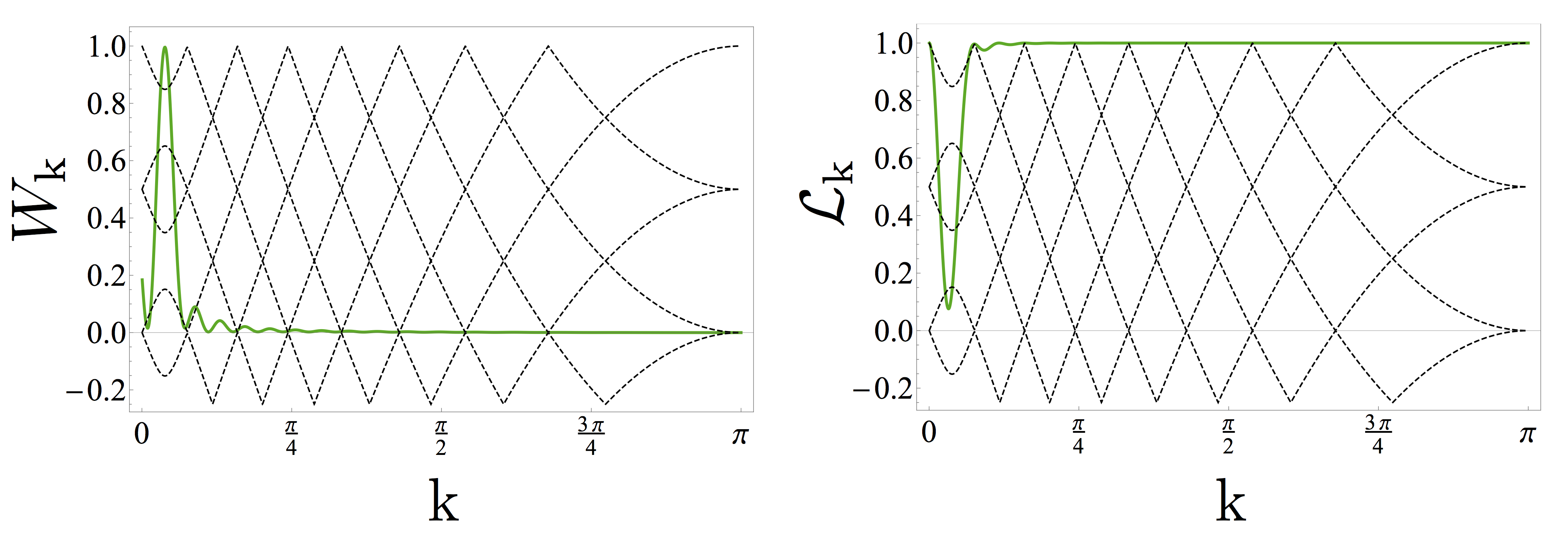}
\caption{{\bf{Low-frequency drive}} (upper panel)Work performed by
the driving agent after $t= T$, resolved in $k$: the resonances in
the Floquet spectrum move towards the low-$k$ region of the
spectrum of the unperturbed Hamiltonian ${\hat{H}}(0)$. (lower
panel)  Momentum resolved Loschmidt echo evaluated at the fixed
time $t= T$, showing that, for the modes near the
quasi-degeneracy, now in the FSS-relevant region, the decay of the
Loschmidt echo is both faster and more significant than for higher
$\omega$ (see Fig.~4 in the main text). In both plots we
considered a driving frequency $\omega=0.5$.  } \label{F.fitb}
\end{center}
\end{figure}

In the main text of the manuscript we have investigated mainly
drivings at frequencies $\omega>4$, where the FSS behaviour is
resilient under the periodic modulation of the magnetic field
$h(t)$. In this subsection we show that, in the low-frequency
limit, the FSS is lost already after the first cycle. To support
this claim, we report in Fig.~\ref{F.fitb} the $k$-resolved work
and Loschmidt echo for $\omega=0.5$. As already stated in
Ref.~\cite{Angelo}, by decreasing the frequency of the drive, the
number of resonances in the Floquet spectrum increases and, more
importantly, they also move towards the low $k$-modes. As a
consequence, the energy injected into the system by the
low-frequency driving is absorbed mainly by the latter and also
the Loschmidt echo of the low-$k$ modes decreases to values
significantly lower than in the $\omega>4$ case. As a result, the
FSS behaviour is not resilient to such low-$\omega$ drives.

%
\section{Acknowledgments}
%
T.J.G.A. is indebted to Alessandro Silva, Rosario Fazio, Angelo Russomanno, and John Goold for insightful discussions. S.L. thanks Ugo Marzolino for useful correspondence.
S.L.,
F.P., G.M.P. and T.J.G.A. acknowledge financial support from the EU
collaborative project QuPRoCs (Grant Agreement 641277). J.M.
acknowledges support from the Alexander von Humboldt Foundation.

\section{Author contributions}
Salvatore Lorenzo, Jamir Marino, Francesco Plastina, G. Massimo Palma, and Tony J. G. Apollaro
conceived the research, discussed the results and
contributed to the final version of the manuscript.
\section{Additional information}
{\bf{Competing financial interests:} }The authors declare no
competing financial interests.



\end{document}